\documentclass[runningheads]{llncs}
\usepackage{graphicx}
\usepackage{hyperref}
\usepackage{mathrsfs}
\usepackage{amsmath}
\usepackage{dsfont}
\usepackage{amstext}
\usepackage{amsfonts}
\begin{document}
\title{Bspline solids manipulation with Mathematica
}
\titlerunning{Bspline solids manipulation with Mathematica}
%
\author{R. Ipanaqu\'{e}\inst{1}\orcidID{0000-0002-3873-6780} \and
R. Velezmoro\inst{1}\orcidID{0000-0002-2582-8264} \and\\
R. T. Urbina\inst{1}\orcidID{0000-0002-8544-3307}}
\authorrunning{R. Ipanaqu\'{e} et al.}
%
\institute{Universidad Nacional de Piura, Urb. Miraflores s/n, Castilla, Piura, Per\'{u}\\
\email{\{ripanaquec,rvelezmorol,rurbinag\}@unp.edu.pe}}
\maketitle              
\begin{abstract}
Bspline solids are used for solid objects modeling in $\mathds{R}^3$. Mathematica incorporates a several commands to manipulate symbolic and graphically Bspline basis functions and to graphically manipulate Bsplines curves and surfaces; however, it does not incorporate any command to the graphical manipulation of Bspline solids. In this paper, we describe a new Mathematica program to compute and plotting the Bspline solids. The output obtained is consistent with Mathematica's notation. The performance of the commands are discussed by using some illustrative examples.

\keywords{Make mesh, Make polyhedrons, Solids, BSpline solids.}
\end{abstract}
\section{Introduction}
\label{sect:introduction}
Currently, Bspline solids are extensively used for solid objects modeling in $\mathds{R}^3$ \cite{ref_article1,ref_proc1,ref_article2,ref_article3}. The most popular and widely used symbolic computation program, \emph{Mathematica} \cite{ref_book2}, incorporates a several commands to manipulate symbolic and graphically Bspline basis functions and to graphically manipulate Bspline curves and surfaces \cite{ref_url1,ref_url2}; however, it does not incorporate any command to the graphical manipulation of Bspline solids.

In this paper, we describe a new Mathematica command: \texttt{BSplineSolid} (which defines a graphics object), to plotting the Bspline solids. Both commands are coded based on the code that displays R. Maeder \cite{ref_book1} and commands that incorporates Mathematica \cite{ref_url1}, for that reason its options supported  come to be the same as those commands in which them are based. The commands have been implemented in Mathematica v.11.0 although later releases are also supported. The output obtained is consistent with Mathematica's notation.

The structure of this paper is as follows: Section 2 provides some mathematical background on Bspline basis, Bspline curves, Bspline surfaces and Bspline solids. Then, Section 3 describes the main standard Mathematica tools for Bsplines manipulation. Finally, Section 4 introduces the new Mathematica commands for manipulating them. The performance of the commands are discussed by using some illustrative examples.
\section{Mathematical Preliminaries}
\label{sect:math-prelim}
An order $d$ Bspline is formed by joining several pieces of polynomials of degree $d-1$ with at most $C^{k-2}$ continuity at the breakpoints. A set of nondescending breaking points $t_{0}\leq t_{1}\leq \ldots\leq t_{m}$ defines a \emph{knot vector}
\[\mathbf{T}=(t_{0},t_{1},\ldots,t_{m})\,,\]
which determines the parametrization of the basis functions.

Given a knot vector $\mathbf{T}$, the associated Bspline basis functions, $N_{i,d}(t)$, are defined as:
\[N_{i,1}(t)=\begin{cases}
1 & \text{for } t_{i}\leq t<t_{i+1}\\
0 & \text{otherwise}\,,
\end{cases}\]
for $d=1$, and
\[N_{i,d}(t)=\frac{t-t_{i}}{t_{i+d-1}-t_{i}}N_{i,d-1}(t)+\frac{t_{i+d}-t}{t_{i+d}-t_{i+1}}N_{i+1,d-1}(t)\,,\]
for $d>1$ and $i=0,1,\ldots,n$.

The Bspline curve is defined by a set of control points $\mathbf{p}_{i}$, $0\leq i\leq n_{1}$ and a knot vector $\mathbf{U}=(u_{0},u_{1},\ldots,u_{n_{1}+d_{1}})$ associated with the parameter $u$. The corresponding Bspline curve is given by
\[\mathbf{r}(u)=\sum _{i=0}^{n_1} \mathbf{p}_{i} N_{i,d_1}(u)\,.\]

A NURBS curve can be represented as
\[\mathbf{r}(u)=\frac{\sum _{i=0}^{n_1} h_{i} \mathbf{p}_{i} N_{i,d_1}(u)}{\sum _{i=0}^{n_1} h_{i} N_{i,d_1}(u)}\]
where $h_{i}>0$ is a weighting factor. If all $h_{i}=1$, the Bspline curve is recovered.

The Bspline surface is a tensor product surface defined by a topologically rectangular set of control points $\mathbf{p}_{ij}$, $0\leq i\leq n_{1}$, $0\leq j\leq n_{2}$ and two knot vectors $\mathbf{U}=(u_{0},u_{1},\ldots,u_{n_{1}+d_{1}})$ and $\mathbf{V}=(v_{0},v_{1},\ldots,v_{n_{2}+d_{2}})$ associated with each parameter $u,v$. The corresponding Bspline surface is given by
\[\mathbf{r}(u,v)=\sum _{i=0}^{n_1} \sum _{j=0}^{n_2} \mathbf{p}_{i j} N_{i,d_1}(u) N_{j,d_2}(v)\,.\]

A NURBS surface can be represented as
\[\mathbf{r}(u,v)=\frac{\sum _{i=0}^{n_1} \sum _{j=0}^{n_2} h_{i j} \mathbf{p}_{i j} N_{i,d_1}(u) N_{j,d_2}(v)}{\sum _{i=0}^{n_1} \sum _{j=0}^{n_2} h_{i j} N_{i,d_1}(u) N_{j,d_2}(v)}\]
where $h_{i j}>0$ is a weighting factor. If all $h_{i j}=1$, the Bspline surface is recovered.

The Bspline solid is a tensor product solid defined by a topologically paralelepepidal set of control points $\mathbf{p}_{ijk}$, $0\leq i\leq n_{1}$, $0\leq j\leq n_{2}$, $0\leq k\leq n_{3}$ and three knot vectors $\mathbf{U}=(u_{0},u_{1},\ldots,u_{n_{1}+d_{1}})$, $\mathbf{V}=(v_{0},v_{1},\ldots,v_{n_{2}+d_{2}})$ and $\mathbf{W}=(w_{0},w_{1},\ldots,w_{n_{3}+d_{3}})$ associated with each parameter $u,v,w$. The corresponding Bspline solid is given by
\[\mathbf{r}(u,v,w)=\sum _{i=0}^{n_1} \sum _{j=0}^{n_2} \sum _{k=0}^{n_3} \mathbf{p}_{i j k} N_{i,d_1}(u) N_{j,d_2}(v) N_{k,d_3}(w)\,.\]

A NURBS solid can be represented as
\[\mathbf{r}(u,v,w)=\frac{\sum _{i=0}^{n_1} \sum _{j=0}^{n_2} \sum _{k=0}^{n_3} h_{i j k} \mathbf{p}_{i j k} N_{i,d_1}(u) N_{j,d_2}(v) N_{k,d_3}(w)}{\sum _{i=0}^{n_1} \sum _{j=0}^{n_2} \sum _{k=0}^{n_3} h_{i j k} N_{i,d_1}(u) N_{j,d_2}(v) N_{k,d_3}(w)}\]
where $h_{i j k}>0$ is a weighting factor. If all $h_{i j k}=1$, the Bspline solid is recovered.

\section{Standard Mathematica tools for Bspline manipulation}

The command
\begin{center}
\texttt{BSplineBasis[}$\{d,\{u_{1},u_{2},\ldots\}\},n,t$\texttt{]}
\end{center}
gives the $n^{\text{th}}$ non-uniform B-spline basis function of degree $d$ with knots at positions $u_{i}$.

\begin{figure}
\centering
\includegraphics[scale=0.375]{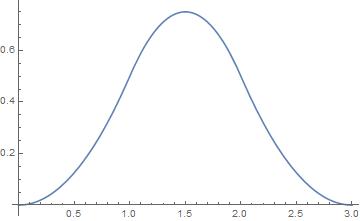}\;
\includegraphics[scale=0.375]{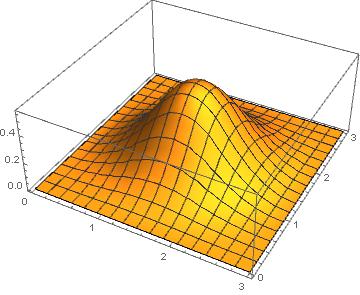}
\caption{Bspline basis functions in $\mathds{R}^{2}$ (left) and $\mathds{R}^{3}$ (right).}
\label{fig:basisf1}
\end{figure}

Below are two examples of base functions, in $\mathds{R}^{2}$ and $\mathds{R}^{3}$, generated with the BsplineBasis command.
\begin{verbatim}
In[1]:=U={0,0,0,1,2,3,3,3};
       Plot[BSplineBasis[{2,U},2,u],{u,0,3}]
\end{verbatim}
$Out[2]=$\emph{See Figure \ref{fig:basisf1}.}

\begin{verbatim}
In[3]:=U=V={0,0,0,1,2,3,3,3};
       Plot3D[BSplineBasis[{2,U},2,u] BSplineBasis[{2,V},2,v],
        {u,0,3},{v,0,3},PlotRange->All]
\end{verbatim}
$Out[4]=$\emph{See Figure \ref{fig:basisf1}.}

The command
\begin{center}
\texttt{BSplineCurve[}$\{pt_{1},pt_{2},\ldots\},options$\texttt{]}
\end{center}
is a graphics primitive that represents a nonuniform rational B-spline curve with control points $pt_{i}$. This command incorporates the options
\begin{center}
\texttt{SplineKnots}, \texttt{SplineWeights}, \texttt{SplineDegree} and \texttt{SplineClosed}
\end{center}
The first three options allow to control the knot vector, the weights and the degree (respectively) that define the Bspline curve. The last option allows defining a closed Bspline curve.

\begin{figure}
\centering
\includegraphics[scale=0.35]{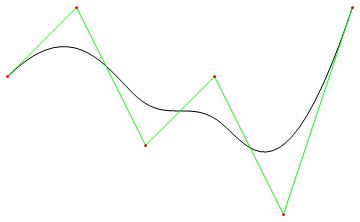}\;
\includegraphics[scale=0.35]{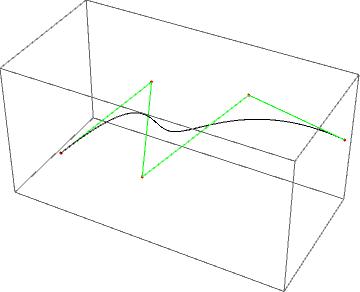}
\caption{Bspline curves in $\mathds{R}^{2}$ (left) and $\mathds{R}^{3}$ (right) and its control points.}
\label{fig:bscurves}
\end{figure}

The two examples shown below correspond to a Bspline curve in $\mathds{R}^{2}$ and another Bspline curve in $\mathds{R}^{3}$.
\begin{verbatim}
In[5]:=pts={{0,0},{1,1},{2,-1},{3,0},{4,-2},{5,1}};
       Graphics[{BSplineCurve[pts],Green,Line[pts],Red,Point[pts]}]
\end{verbatim}
$Out[6]=$\emph{See Figure \ref{fig:bscurves}.}
\begin{verbatim}
In[7]:=pts={{0,0,0},{1,1,1},{2,-1,1},{3,0,2},{4,1,1}};
       Graphics3D[{BSplineCurve[pts],Green,Line[pts],Red,Point[pts]}]
\end{verbatim}
$Out[8]=$\emph{See Figure \ref{fig:bscurves}.}\medskip

The command
\begin{center}
\texttt{BSplineSurface[}$array,options$\texttt{]}
\end{center}
is a graphics primitive that represents a nonuniform rational B-spline surface defined by an array of $x,y,z$ control points. This command incorporates the same options as the \texttt{BsplineCurve} command.

\begin{figure}
\centering
\includegraphics[scale=0.425]{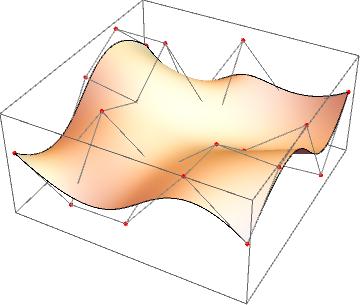}
\caption{Control points together with the Bspline surface.}
\label{fig:bssurfaces}
\end{figure}

The following example shows a Bspline surface for an random array of control points.
\begin{verbatim}
In[9]:=cpts=Table[{i,j,RandomReal[{-1,1}]},{i,5},{j,5}];
       Graphics3D[{BSplineSurface[cpts],
        PointSize[Medium], Red, Map[Point, cpts], Gray, Line[cpts], 
        Line[Transpose[cpts]]}]
\end{verbatim}
$Out[10]=$\emph{See Figure \ref{fig:bssurfaces}.}\medskip

\section{Bspline solids manipulation}

We will start this section by defining the \texttt{MakeMesh} \cite{ref_book1} and \texttt{MakePolyhedrons} (new) commands to build a flat Bspline surface (entirely contained in a 2D space) and a spatial Bspline solid (entirely contained in a 3D space), respectively.

\begin{verbatim}
In[10]:=MakeMesh[vl_List]:=
         Block[{l=vl,l1=Map[RotateLeft,vl],mesh},
           mesh={l,l1,RotateLeft[l1],RotateLeft[l]};
           mesh=Map[Drop[#,-1]&,mesh,{1}];
           mesh=Map[Drop[#,-1]&,mesh,{2}];
           Transpose[Map[Flatten[#,1]&,mesh]]
           ]
\end{verbatim}

\begin{verbatim}
In[12]:=MakePolyhedrons[vl_List]:=
         Block[{ mesh3d,newfaces,
             faces = {{1,2,3,4},{5,6,7,8},{1,2,6,5},
             {8,7,3,4},{2,3,7,6},{1,5,8,4}} },
           mesh3d=Map[MakeMesh,vl];
           mesh3d=
            Map[Flatten[#,1]&,Map[Thread,Partition[mesh3d,2,1]],{2}];
           mesh3d=Flatten[mesh3d,2];
           newfaces=Table[faces+8(i-1),{i,Length[mesh3d]/8}];
           GraphicsComplex[mesh3d,GraphicsGroup[Polygon/@newfaces]]
           ]
\end{verbatim}

\begin{figure}
\centering
\includegraphics[scale=0.3]{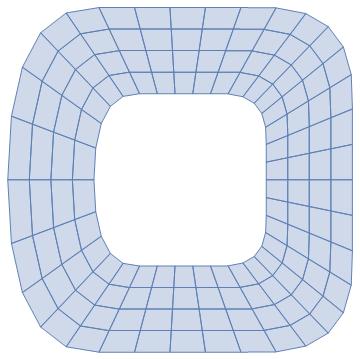}\;
\includegraphics[scale=0.4,trim=0 20 0 0,clip]{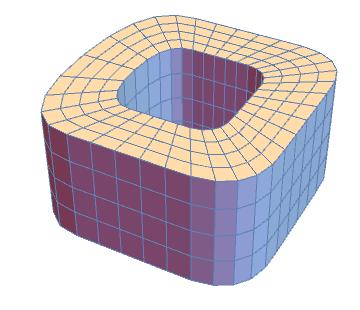}
\caption{A flat Bspline surface (left) and a spatial Bspline solid (right).}
\label{fig:princip}
\end{figure}

Next, the control points, the node vectors and the weights defining a flat Bspline surface are provided. Additionally, such a surface is plotted (Fig. \ref{fig:princip}). All this is done using the built-in Mathematica commands and the \texttt{MakeMesh} command.

\begin{verbatim}
In[13]:=points=
         {{{-2,0},{-3/2,0},{-1,0}},{{-2,-2},{-3/2,-3/2},{-1,-1}},
         {{0,-2},{0,-3/2},{0,-1}},{{2,-2},{3/2,-3/2},{1,-1}},
         {{2,0},{3/2,0},{1,0}},{{2,2},{3/2,3/2},{1,1}},
         {{0,2},{0,3/2},{0,1}},{{-2,2},{-3/2,3/2},{-1,1}},
         {{-2,0},{-3/2,0},{-1,0}}};
\end{verbatim}

\begin{verbatim}
In[14]:={pp1,pp2}={35,5};
        {d1,d2}={2,2};
        {knots1,knots2}={{0,0,0,1,2,3,4,5,6,7,7,7},{0,0,0,1,1,1}};
        sw=Table[1,{i,n1},{j,n2}];
\end{verbatim}

\begin{verbatim}
In[18]:=bssurf=Sum[sw[[i+1,j+1]]points[[i+1,j+1]]
         BSplineBasis[{d1,knots1},i,u]BSplineBasis[{d2,knots2},j,v],
         {i,0,n1-1},{j,0,n2-1}]/
         Sum[sw[[i+1,j+1]]
         BSplineBasis[{d1,knots1},i,u]BSplineBasis[{d2, knots2},j,v],
         {i,0,n1-1},{j,0,n2-1}];
\end{verbatim}

\begin{verbatim}
In[19]:={u0,u1,v0,v1}={knots1[[d1+1]],knots1[[n1+d1 -1]],
          knots2[[d2+1]],knots2[[n2+d2-1]]};
        {du,dv}={(u1-u0)/(pp1-1.),(v1-v0)/(pp2-1.)};
        aux=Table[N[bssurf],{u,u0,u1,du},{v,v0,v1,dv}];
\end{verbatim}

\begin{verbatim}
In[22]:=Graphics[{EdgeForm[RGBColor[0.36841,0.50677,0.70979]], 
         Directive[RGBColor[0.36841,0.50677,0.70979],Opacity[0.3]], 
         Polygon/@MakeMesh[aux]}]
\end{verbatim}
$Out[22]=$\emph{See Figure \ref{fig:princip} (left).}\medskip

Then, control points, node vectors and weights that define a spatial Bspline solid are provided. In addition, such a solid is plotted (Fig. \ref{fig:princip}). All this is done using the built-in Mathematica commands and the \texttt{MakePolyhedrons} command.

\begin{verbatim}
In[23]:=points={{{{-2,0,0},{-3/2,0,0},{-1,0,0}},{{-2,-2,0},
            {-3/2,-3/2,0},{-1,-1,0}},{{0,-2,0},{0,-3/2,0},
            {0,-1,0}},{{2,-2,0},{3/2,-3/2,0},{1,-1,0}},
           {{2,0,0},{3/2,0,0},{1,0,0}},{{2,2,0},{3/2,3/2,0},
            {1,1,0}},{{0,2,0},{0,3/2,0},{0,1,0}},{{-2,2,0},
            {-3/2,3/2,0},{-1,1,0}},{{-2,0,0},{-3/2,0,0},{-1,0,0}}},
         {{{-2,0,1},{-3/2,0,1},{-1,0,1}},{{-2,-2,1},{-3/2,-3/2,1},
            {-1,-1,1}},{{0,-2,1},{0,-3/2,1},{0,-1,1}},{{2,-2,1},
            {3/2,-3/2,1},{1,-1,1}},{{2,0,1},{3/2,0,1},{1,0,1}},
           {{2,2,1},{3/2,3/2,1},{1,1,1}},{{0,2,1},{0,3/2,1},
            {0,1,1}},{{-2,2,1},{-3/2,3/2,1},{-1,1,1}},{{-2,0,1},
            {-3/2,0,1},{-1,0,1}}},{{{-2,0,2},{-3/2,0,2},{-1,0,2}},
           {{-2,-2,2},{-3/2,-3/2,2},{-1,-1,2}},{{0,-2,2},{0,-3/2,2},
            {0,-1,2}},{{2,-2,2},{3/2,-3/2,2},{1,-1,2}},{{2,0,2},
            {3/2,0,2},{1,0,2}},{{2,2,2},{3/2,3/2,2},{1,1,2}},
           {{0,2,2},{0,3/2,2},{0,1,2}},{{-2,2,2},{-3/2,3/2,2},
            {-1,1,2}},{{-2,0,2},{-3/2,0,2},{-1,0,2}}}};
\end{verbatim}

\begin{verbatim}
In[24]:={n1,n2,n3,d}=Dimensions[points];
        {pp1,pp2,pp3}={5,35,5};
        {d1,d2,d3}={2,2,2};
        {knots1,knots2,knots3}={{0,0,0,1,1,1},
          {0,0,0,1,2,3,4,5,6,7,7,7},{0,0,0,1,1,1}};
        sw = Table[1,{i,n1},{j,n2},{k,n3}];
\end{verbatim}

\begin{verbatim}
In[29]:=bssol=Sum[sw[[i+1,j+1,k+1]]points[[i+1,j+1,k+1]]
          BSplineBasis[{d1,knots1},i,u]BSplineBasis[{d2,knots2},j,v]
          BSplineBasis[{d3,knots3},k,w],
           {i,0,n1-1},{j,0,n2-1},{k,0,n3-1}]/
         Sum[sw[[i+1,j+1,k+1]]BSplineBasis[{d1,knots1},i,u]
          BSplineBasis[{d2,knots2},j,v]BSplineBasis[{d3,knots3},k, w],
          {i,0,n1-1},{j,0,n2-1},{k,0,n3-1}];
\end{verbatim}

\begin{verbatim}
In[30]:={u0,u1,v0,v1,w0,w1}={knots1[[d1+1]],knots1[[n1+d1-1]],
         knots2[[d2+1]],knots2[[n2+d2-1]],knots3[[d3+1]],
         knots3[[n3+d3-1]]};
        {du,dv,dw}={(u1-u0)/(pp1-1.),(v1-v0)/(pp2-1.),
         (w1-w0)/(pp3-1.)};
        aux=Table[N[bssol],{u,u0,u1,du},{v,v0,v1,dv},{w,w0,w1,dw}];
\end{verbatim}

\begin{verbatim}
In[33]:=Graphics3D[{EdgeForm[RGBColor[0.36841,0.50677,0.70979]],
         MakePolyhedrons[aux]},Boxed->False]
\end{verbatim}
$Out[33]=$\emph{See Figure \ref{fig:princip} (right).}\medskip

By comparing the two previous processes it is possible to form an idea of the algorithm used to implement the main command, \texttt{BSplineSolid}.

Finally, we describe the Mathematica program we developed to compute the Bspline solids. For the sake of clarity, the program will be explained through some illustrative examples.

The \texttt{BSplineSolid} command is a graphics primitive that represents a nonuniform rational Bspline solid defined by an array of $x,y,z$ control points. The syntax of this command is:
\[\texttt{BSplineSolid[}pts,options\texttt{]}\]

The options supported by this command are: \texttt{PlotPoints}, \texttt{SplineDegree}, \texttt{SplineKnots} and \texttt{SplineWeights}.

\begin{figure}
\centering
\includegraphics[scale=0.355]{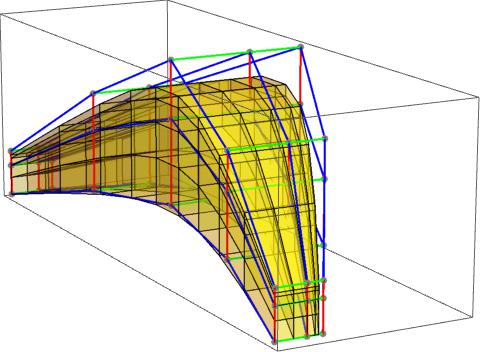}\;
\includegraphics[scale=0.35]{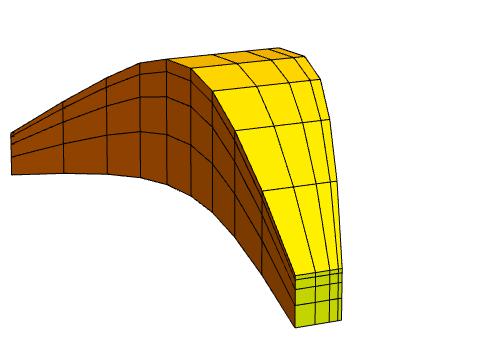}
\caption{Bspline solid.}
\label{fig:solid2}
\end{figure}

As a first example we compute the Bspline solid of order 2 for a set of points as follows:
\begin{verbatim}
In[34]:=pts=Table[{u,1/(1+u^2+v^2),1/(1+u^2+w^2)},
         {u,-1,1,1/2},{v,-1,1,1/2},{w,-1,1,1/2}];
\end{verbatim}

\begin{verbatim}
In[35]:=Graphics3D[{Meshing[pts],Yellow,Opacity[0.2],BSplineSolid[pts]},
         ViewPoint->{3.,-1.6,0.8}]
\end{verbatim}
$Out[35]=$\emph{See Figure \ref{fig:solid2} (left).}\medskip

The \texttt{Meshing} command is defined as follows:
\begin{verbatim}
In[36]:=Meshing[pts_?ArrayQ]:={AbsolutePointSize[7],Gray, 
         Map[Point,pts,{2}],Thick,Red,Map[Line[#] &,pts,{2}],
         Green,Map[Line[#]&,Transpose/@pts,{2}],
         Blue,Map[Line[#]&,Transpose/@Transpose[Transpose/@pts],{2}]}
\end{verbatim}

\begin{verbatim}
In[37]:=Graphics3D[{Yellow,BSplineSolid[pts]},
         ViewPoint->{3.,-1.6,0.8},Boxed->False]
\end{verbatim}
$Out[37]=$\emph{See Figure \ref{fig:solid2} (right).}\medskip

\begin{figure}
\centering
\includegraphics[scale=0.475]{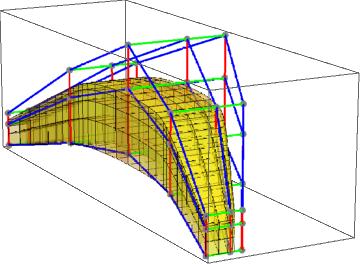}\;
\includegraphics[scale=0.425]{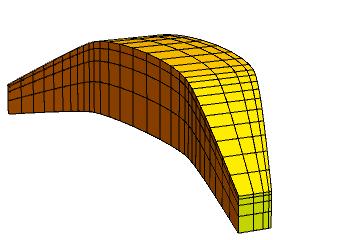}
\caption{NURBS solid.}
\label{fig:solid3}
\end{figure}

As a second example we compute the NURBS solid of order 2 for a set of points as follows:
\begin{verbatim}
In[38]:=w1={1,1,0.25,1,1};
        w3=w2=w1;
        w = Table[w1[[i]]w2[[j]]w3[[k]],{i,1,5},{j,1,5},{k,1,5}];
\end{verbatim}

\begin{verbatim}
In[41]:=Graphics3D[{Meshing[pts],Yellow,Opacity[0.2],
         BSplineSolid[pts,SplineWeights->w,PlotPoints->{30,10,10}]},
         ViewPoint->{3.,-1.6,0.8}]
\end{verbatim}
$Out[41]=$\emph{See Figure \ref{fig:solid3} (left).}\medskip

\begin{verbatim}
In[42]:=Graphics3D[{Yellow,BSplineSolid[pts,SplineWeights->w,
          PlotPoints->{30,10,10}]},ViewPoint->{3.,-1.6,0.8},
         Boxed->False]
\end{verbatim}
$Out[42]=$\emph{See Figure \ref{fig:solid3} (right).}\medskip

\begin{figure}
\centering
\includegraphics[scale=0.325]{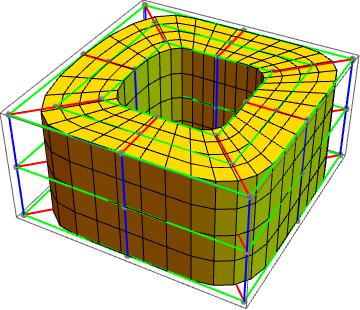}\qquad
\includegraphics[scale=0.325]{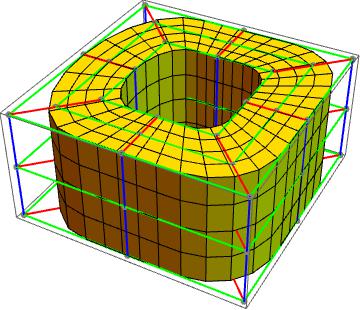}\\
\includegraphics[scale=0.325]{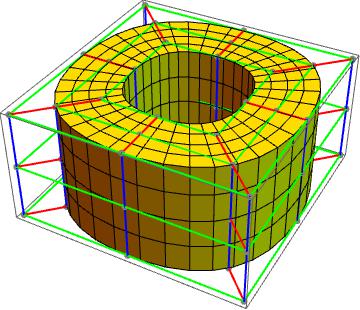}\qquad
\includegraphics[scale=0.325]{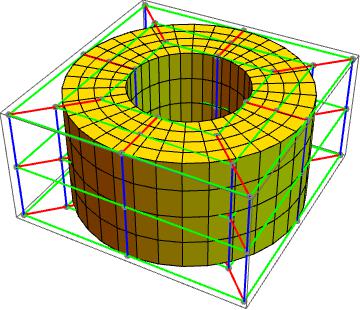}
\caption{Modeling a hollow cylinder with NURBS solid.}
\label{fig:hollowcyl}
\end{figure}

\begin{figure}
\centering
\includegraphics[scale=0.325]{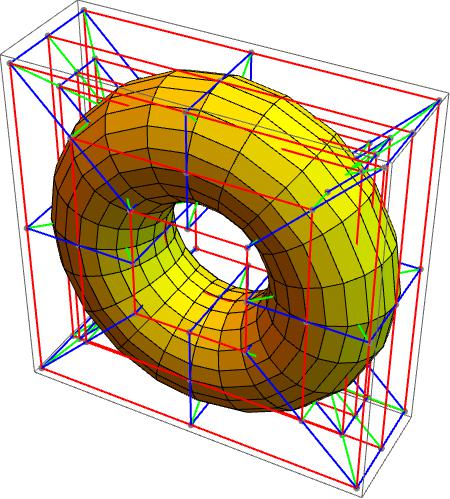}\qquad
\includegraphics[scale=0.355]{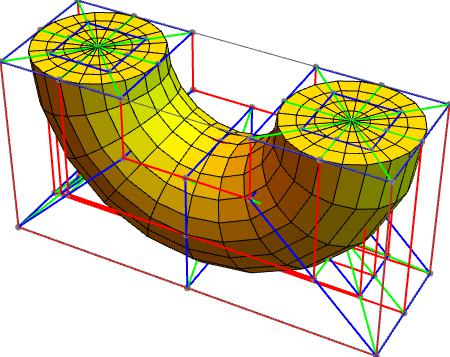}
\caption{Modeling a solid torus with NURBS solid.}
\label{fig:torus}
\end{figure}

\begin{figure}
\centering
\includegraphics[scale=0.325]{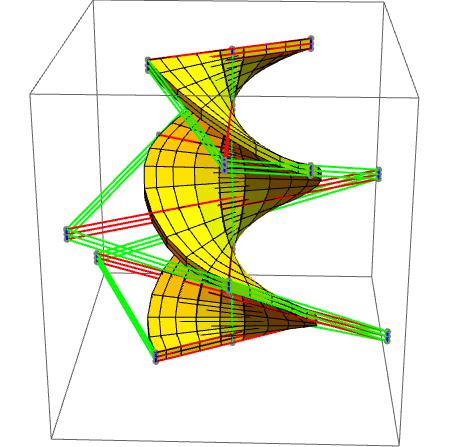}\qquad
\includegraphics[scale=0.325]{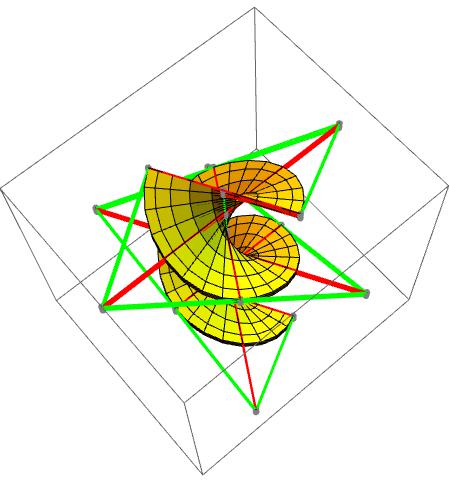}
\caption{Modeling a solid helix with NURBS solid.}
\label{fig:helix}
\end{figure}

Moreover, figure \ref{fig:hollowcyl} shows the adjustment process, by varying the node vectors and weights, modeling a solid hollow cylinder with NURBS solid. In practice this is achieved by changing the default values of options \texttt{SplineWeights} and \texttt{SplineKnots}. Similarly, seen in figure \ref{fig:torus} the modeling a solid torus and in figure \ref{fig:helix} the modeling a solid helix.

\section{Conclusions}

In this paper, we describe a new Mathematica program to compute and plotting the Bspline solids. The output obtained is consistent with Mathematica's notation. The performance of the commands are discussed by using some illustrative examples.

\section{Acknowledgements}

The authors would like to thank to the authorities of the Universidad Nacional de Piura for the acquisition of the Mathematica 11.0 license and the reviewers for their valuable comments and suggestions.
%
%
%
%

\end{document}